\documentclass{article}

\usepackage[preprint]{spconf}
\usepackage{graphicx}

\usepackage{color}
\usepackage{amsmath,amssymb,amsfonts,dsfont}
\usepackage{multirow}
\usepackage{subfigure}
\usepackage{booktabs} % For thickline in tables
\usepackage[table]{xcolor}

\usepackage{tikz}
\usepackage{pgfplots}
\pgfplotsset{width=10cm,compat=1.15}
\usepgfplotslibrary{statistics}
\usepgfplotslibrary{colormaps}
\usepgfplotslibrary{fillbetween}

\usepackage{url}
\usepackage[hidelinks]{hyperref}
\hypersetup{
    colorlinks=true,
    linkcolor=black,
    filecolor=black,
    urlcolor=gray,
    citecolor=black,
    breaklinks=true
}

\newcommand{\bbox}{\protect\raisebox{1pt}{\protect\tikz \protect\draw[black,fill=black] (1,1) circle (0.5ex);}}
\newcommand{\wbox}{\protect\raisebox{1pt}{\protect\tikz \protect\draw[black,fill=white] (1,1) circle (0.5ex);}}

\definecolor{mylightgray}{gray}{0.9}

\title{AUDIO-VISUAL OBJECT CLASSIFICATION FOR HUMAN-ROBOT COLLABORATION}

\name{A. Xompero$^{1}$, Y. L. Pang$^{1}$, T. Patten$^{2}$, A. Prabhakar$^{3}$, B. Calli$^{4}$, A. Cavallaro$^{1}$\thanks{This work is supported by the CHIST-ERA program through the project CORSMAL, under UK EPSRC grant EP/S031715/1 and Swiss NSF grant 20CH21{\_}180444.}}
\address{$^{1}$Queen Mary University of London, UK, $^{2}$University of Technology Sydney, Australia,\\$^{3}$École polytechnique fédérale de Lausanne, Switzerland, $^{4}$Worcester Polytechnic Institute, USA}

\begin{document}

\setcounter{page}{0}
\twocolumn[\noindent Copyright 2022 IEEE. Published in ICASSP 2022 - 2022 IEEE International Conference on Acoustics, Speech and Signal Processing (ICASSP), scheduled for 7-13 May 2022 Virtual; 22-27 May 2022 In-Person in Singapore. Personal use of this material is permitted. However, permission to reprint/republish this material for advertising or promotional purposes or for creating new collective works for resale or redistribution to servers or lists, or to reuse any copyrighted component of this work in other works, must be obtained from the IEEE. Contact: Manager, Copyrights and Permissions / IEEE Service Center / 445 Hoes Lane / P.O. Box 1331 / Piscataway, NJ 08855-1331, USA. Telephone: + Intl. 908-562-3966.]

\ninept

\maketitle
%
% \copyrightnotice{\copyright\ 2022 IEEE.}
% \toappear{To appear in {\it Proc.\ ICASSP2022,
                    % May 22-27, 2022, Singapore}}

\begin{abstract}
Human-robot collaboration requires the contactless estimation of the physical properties of containers manipulated by a person, for example while pouring content in a cup or moving a food box. Acoustic and visual signals can be used to estimate the physical properties of such objects, which may vary substantially in shape, material and size, and also be occluded by the hands of the person. To facilitate comparisons and stimulate progress in solving this problem, we present the CORSMAL challenge and a dataset to assess the performance of the algorithms through a set of well-defined performance scores. The tasks of the challenge are the estimation of the mass, capacity, and dimensions of the object (container), and the classification of the type and amount of its content. A novel feature of the challenge is our real-to-simulation framework for visualising and assessing the impact of estimation errors in human-to-robot handovers.
\end{abstract}

\section{Introduction}
\label{sec:intro}

Robots supporting people in their daily activities at home or at the workplace need to accurately and robustly perceive objects, such as containers, and their physical properties, for example when they are manipulated by a person prior to a human-to-robot handover~\cite{Sanchez-Matilla2020,Medina2016,Rosenberger2021RAL,Ortenzi2021TRO,Yang2021ICRA}. Audio-visual perception should adapt -- on-the-fly and with limited or no prior knowledge -- to changing conditions in order to guarantee the correct execution of the task and the safety of the person. For assistive scenarios at home, audio-visual perception should accurately and robustly estimate the physical properties (e.g., weight and shape) of household containers, such as cups, drinking glasses, mugs, bottles, and food boxes~\cite{Sanchez-Matilla2020,Ortenzi2021TRO,Liang2020MultimodalPouring,Modas2021ArXiv,Xompero2021_ArXiv}. However, the material, texture, transparency and shape can vary considerably across containers and also change with their content, which may not be visible due to the opaqueness of the container or occlusions, and hence should be inferred through the behaviour of the human~\cite{Sanchez-Matilla2020,Modas2021ArXiv,Xompero2021_ArXiv,Mottaghi2017ICCV,Duarte2020ICDL_EpiRob}. 

\begin{figure}[t!]
    \centering
    \includegraphics[width=\linewidth]{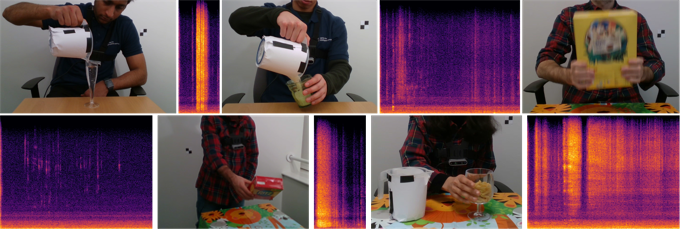}
    \caption{Sample video frames and audio spectrograms of people manipulating objects prior to handing them over to a robot.}
    \label{fig:avsamples}
\end{figure}

\begin{figure*}[t!]
    \centering
    \includegraphics[width=\textwidth]{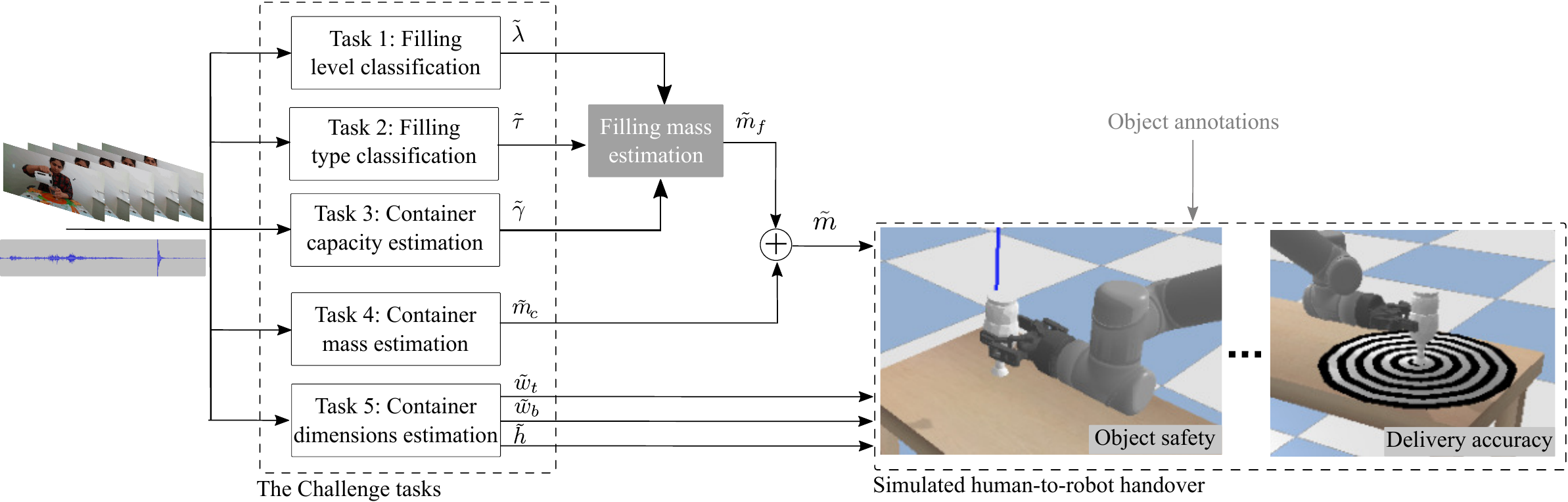}
    \caption{The challenge tasks feeding into the CORSMAL simulator~\cite{Pang2021ROMAN} to evaluate the impact of estimation errors. Given video frames and audio signals from the CORSMAL Containers Manipulation (CCM) dataset~\cite{Xompero2021_ArXiv,Xompero_CCM}),  the results of T1 (filling level), T2 (filling type), and T3 (container capacity) are used to compute the filling mass, which is added to T4 (container mass) for estimating the  mass of the object (container + filling). The estimated dimensions (T5) are used to visualise the container. The simulator also uses object annotations, such as 6D poses over time, the true weight (container + filling), a 3D mesh model reconstructed offline with a vision baseline~\cite{Pang2021ROMAN}, and the frame where the object is ready to be grasped by the simulated robot arm, for performing and visualising the handover. 
    }
    \label{fig:challengetasksdiagram}
\end{figure*}

In this paper, we present the tasks and the results of the CORSMAL challenge at IEEE ICASSP 2022, supporting the design and evaluation of audio-visual solutions for the estimation of the physical properties of a range of containers manipulated by a person prior to a handover (see Fig.~\ref{fig:avsamples}). 
The specific containers and fillings are not known in advance, and the only priors are the sets of object categories ({drinking glasses}, {cups}, {food boxes}) and filling types ({water}, {pasta}, {rice}). The estimation of the mass and dimensions of the containers are novel tasks of this challenge, and complement the tasks of its previous version~\cite{Xompero2021_ArXiv}, such as the estimation of the container capacity and the type, mass and amount of the content. We carefully defined a set of performance scores to directly evaluate and systematically compare the algorithms on each task. Moreover, to assess the accuracy of the estimations and visualise the safeness of human-to-robot handovers, we implemented a real-to-simulation framework~\cite{Pang2021ROMAN} that provides indirect high-level evaluations on the impact of these tasks (see Fig.~\ref{fig:challengetasksdiagram}). The source code of the entries to the challenge and the up-to-date leaderboards are available at 
\mbox{\url{http://corsmal.eecs.qmul.ac.uk/challenge.html}}.

\section{The tasks}
\label{sec:tasks}

In the scope of the challenge and based on the reference dataset~\cite{Xompero2021_ArXiv,Xompero_CCM}, containers vary in shape and size, and may be empty or filled with an unknown content at 50\% or 90\% of its capacity. 
We define a configuration as the manipulation of a container with a filling type and amount under a specific setting (i.e., background, illumination, scenario). 
The challenge features five tasks (Ts), each associated with a physical property to estimate for each configuration $j$.
\begin{description}
    \item[Filling level classification (T1).] The goal is to classify the filling level ($\tilde{\lambda}^j$) as empty, 50\%, or 90\%. 
    \item[Filling type classification (T2).] The goal is to classify the type of filling ($\tilde{\tau}^j$), if any, as one of these classes: 0 (no content), 1 (pasta), 2 (rice), 3 (water). 
    \item[Container capacity estimation (T3).] The goal is to estimate the capacity of the container ($\tilde{\gamma}^j$, in mL).
    \item[Container mass estimation (T4).] The goal is to estimate the mass of the (empty) container ($\tilde{m}_{c}^j$, in g).
    \item[Container dimensions estimation (T5).] The goal is to estimate the width at the top ($\tilde{w}_t^j$, in mm) and at the bottom ($\tilde{w}_b^j$, in mm), and height ($\tilde{h}^j$, in mm) of the container. 
\end{description}

Algorithms designed for the challenge are expected to estimate these physical properties to compute the mass of the filling as 
\begin{equation}
    \tilde{m}_f^j = \tilde{\lambda}^j \tilde{\gamma}^j D(\tilde{\tau}^j),
    \label{eq:fillingmass}
\end{equation}
where $D(\cdot)$ selects a pre-computed density based on the classified filling type. The mass of the object $\tilde{m}$ is calculated as the sum of the mass of the empty container and the mass of the content, if any.

\section{The evaluation}
\label{sec:evaluation}

\subsection{Data}

CORSMAL Containers Manipulation (CCM)~\cite{Xompero2021_ArXiv,Xompero_CCM} is the reference dataset for the challenge and  consists of 1,140 visual-audio-inertial recordings of people interacting with 15 container types: 5 drinking cups, 5 drinking glasses, and 5 food boxes.
These containers are made of different materials, such as plastic, glass, and cardboard. Each container can be empty or filled with water, rice or pasta at two different levels of fullness: 50\% or 90\% with respect to the capacity of the container. In total, 12 subjects of different gender and ethnicity\footnote{An individual who performs the manipulation is referred to as \textit{subject}. Ethical approval (QMREC2344a) was obtained at Queen Mary University of London, and consent from each person was collected prior to data collection.} were invited to execute a set of 95 configurations as a result of the combination of containers and fillings, and for one of three manipulation scenarios. The scenarios are designed with an increasing level of difficulty caused by occlusions or subject motions, and recorded with two different backgrounds and two different lighting conditions to increase the visual challenges for the algorithms. The annotation of the data includes the capacity, mass, maximum width  and height (and depth for boxes) of each container, and the type, level, and mass of the filling. The density of pasta and rice is computed from the annotation of the filling mass, capacity of the container, and filling level for each container. Density of water is 1 g/mL. 
For validation, CCM is split into a training set (recordings of 9 containers), a public test set (recordings of 3 containers), and a private test set (recordings of 3 containers). The containers for each set are evenly distributed among the three categories. 
The annotations are provided publicly only for the training set.

\subsection{Real-to-sim visualisation}

The challenge adopts a real-to-simulation framework~\cite{Pang2021ROMAN} that complements the CCM dataset with a human-to-robot handover in the PyBullet simulation environment~\cite{coumans2019pybullet}. The framework uses the physical properties of a manipulated container estimated by a perception algorithm.
The handover setup recreated in simulation consists of a 6 DoF robotic arm (UR5) equipped with a 2-finger parallel gripper (Robotiq 2F-85), and two tables.

The simulator renders a 3D object model reconstructed offline by a vision baseline in manually selected frames with no occlusions~\cite{Pang2021ROMAN}. The weight of the object used by the simulator is the true, annotated value. We manually annotated the poses of the containers for each configuration of CCM every 10 frames and interpolated the intermediate frames. We also annotated the frame where the person started delivering the object to the robot arm. We use the annotated and interpolated poses to render the motion of the object in simulation and control the robot arm to approach the object at the annotated frame for the handover. If the robot is not able to reach the container before the recording ends, the last location of the container is kept for 2~s.

When reaching the container, the simulated robot arm closes the gripper to 2~cm less than the object width to ensure good contact with the object, and applies an amount of force determined by the estimated weight of the object to grasp the container. Note that in the scope of the challenge, we avoid simulating the human hands so that the object is fully visible and can be grasped by the robot arm. The simulator visualises whether the estimations enable the robot to successfully grasp the container without dropping it or squeezing it. After grasping the container, the robot  delivers it to a target area on a table via a predefined trajectory.

\subsection{Scores}

To provide sufficient granularity into the behaviour of the various components of the audio-visual algorithms and pipelines, we compute {13 performance scores} individually for the public test set (no annotations available to the participants), the private test set (neither data nor annotations are available to the participants), and their combination. All scores are in the range $[0,1]$. With reference to Table~\ref{tab:scores}, the first 7 scores quantify the accuracy of the estimations for the 5 main tasks and include filling level, filling type, container capacity, container width at the top, width at the bottom, and height, and container mass. 
Other 3 scores evaluate groups of tasks and assess filling mass, joint filling type and level classification, joint container capacity and dimensions estimation. The last 2 scores are an indirect evaluation of the impact of the estimations (i.e., the object mass) on the quality of human-to-robot handover and delivery of the container by the robot in simulation.

\textbf{T1 and T2.} For filling level and type classification, we compute precision, recall, and F1-score for
each class $k$ across all the configurations of that class, $J_k$. \textit{Precision} is the number of true positives divided by the total number of true positives and false positives for each class $k$ ($P_k$). \textit{Recall} is the number of true positives divided by the total number of true positives and false negatives for each class $k$ ($R_k$). \textit{F1-score} is the harmonic mean of precision and recall, defined as
\begin{equation}
    F_k = 2\frac{P_k R_k}{P_k + R_k}.
\end{equation}
We compute the weighted average F1-score across $K$ classes as, 
\begin{equation}
    \bar{F}_1 = \sum_{k=1}^K \frac{J_k F_k}{J},
    \label{eq:wafs}
\end{equation}
where $J$ is the total number of configurations (for either the public test set, the private test set, or their combination). Note that $K=3$ for the task of filling level classification and $K=4$ for the task of filling type classification.

\textbf{T3, T4 and T5.} For container capacity and mass estimation, we compute the relative absolute error between the estimated measure, $a \in \{\tilde{\gamma}^j, \tilde{m}_c^j \}$, and the true measure, $b \in \{\gamma^j, m_c^j \}$: 
\begin{equation}
    \varepsilon(a, b) = \frac{|a - b |}{b}.
    \label{eq:ware}
\end{equation}

For container dimensions estimation, where $a \in \left\{\tilde{w}_t^j,\tilde{w}_b^j,\tilde{h}^j\right\}$ and $b$ is the corresponding annotation, we use the normalisation function $\sigma_1(\cdot,\cdot)$~\cite{Sanchez-Matilla2020}:
\begin{equation}
 \sigma_1(a,b)=
    \begin{cases}
     1 - \frac{|a - b |}{b} & \text{if} \quad | a - b | < b,  \\
     0  & \text{otherwise}.
    \end{cases}   
\end{equation}

For filling mass estimation\footnote{Note that an algorithm with lower scores for T1, T2 and T3, may obtain a higher filling mass score than other algorithms due to the multiplicative formula to compute the filling mass for each configuration.}, we compute the relative absolute error between the estimated, $\tilde{m}_{f}^j$, and the true filling mass, $m_{f}^j$, unless the annotated mass is zero (empty filling level),
\begin{equation}
    \epsilon(\tilde{m}_f^j, m_f^j) = 
    \begin{cases}
    0, & \text{if } m_f^j = 0 \land \tilde{m}_f^j=0, \\
    \tilde{m}_f^j & \text{if } m_f^j = 0 \land \tilde{m}_f^j \neq 0, \\
    \frac{|\tilde{m}_f^j - m_f^j |}{m_f^j} & \text{otherwise}.
    \end{cases}
    \label{eq:ware2}
\end{equation}

With reference to Table~\ref{tab:scores}, we compute the score, $s_i$, with \mbox{$i=\left\{3,\dots,8\right\}$}, across all the configurations $J$ for each measure as:
\begin{equation}
\noindent
    s_i = 
    \begin{cases}
        \frac{1}{J}\sum_{j=1}^{J}{ \mathds{1}_j e^{-\varepsilon(a, b)}} & \text{if} \, a \in \left\{\tilde{\gamma}^j, \tilde{m}_c^j\right\},\\
        \frac{1}{J}\sum_{j=1}^{J}{\mathds{1}_j \sigma_1(a,b)} & \text{if} \, a \in \left\{\tilde{w}^j,\tilde{w}_b^j,\tilde{h}^j\right\},\\
        \frac{1}{J}\sum_{j=1}^{J}{ \mathds{1}_j e^{-\epsilon(a, b)}} & \text{if} \, a=\tilde{m}_f^j.\\
    \end{cases}
\end{equation}
The value of the indicator function, \mbox{$\mathds{1}_j \in \{0,1\}$}, is 0 only when \mbox{$a \in \left\{ \tilde{\gamma}^j, \tilde{m}_c^j, \tilde{w}_t^j,\tilde{w}_b^j,\tilde{h}^j, \tilde{m}_f^j \right\}$} is not estimated in configuration $j$. Note that estimated and annotated measures are strictly positive, $a>0$ and  $b>0$, except for filling mass in the empty case (i.e., $\tilde{\lambda}^j = 0$ or $\tilde{\tau}^j = 0$).

% ===========================
\begin{table*}[t!]
\centering
\scriptsize
\renewcommand{\arraystretch}{1.2}
\setlength\tabcolsep{1.3pt}
\caption{Results of the CORSMAL challenge entries on the combination of the public and private CCM test sets~\cite{Xompero2021_ArXiv,Xompero_CCM}. For a measure $a$, its corresponding ground-truth value is  $\hat{a}$. All scores are normalised and presented in percentages. $\bar{F}_1(\cdot)$ is the weighted average F1-score. Filling amount and type are sets of classes (no unit).
}
\begin{tabular}{ccccclllllccrrrrrrrrr}
\specialrule{1.2pt}{3pt}{0.6pt}
 T1 & T2 & T3 & T4 & T5 & Description &  Unit & Measure & Score & Weight & Type & R2S & RAN & AVG & \cite{Donaher2021EUSIPCO_ACC} & \cite{Liu2020ICPR} & \cite{Ishikawa2020ICPR} & \cite{Iashin2020ICPR} & \cite{Apicella_GC_ICASSP22} & \cite{Matsubara_GC_ICASSP22} & \cite{Wang_GC_ICASSP22}  \\
\specialrule{1.2pt}{3pt}{1pt}
\bbox & \wbox & \wbox & \wbox & \wbox & Filling level &  & $\lambda^j$  & $s_1 = \bar{F}_1(\lambda^1, \ldots, \lambda^J, \hat{\lambda}^1, \ldots, \hat{\lambda}^J)$ & $\pi_1 = 1/8$ & D & \wbox & 37.62 & 33.15 & \textbf{80.84} & 43.53 & 78.56 & 79.65 & \multicolumn{1}{c}{--} & 65.73 & 77.40 \\
\wbox & \bbox & \wbox & \wbox & \wbox & Filling type &  & $\tau^j$  & $s_2 = \bar{F}_1(\tau^1, \ldots, \tau^J, \hat{\tau}^1, \ldots, \hat{\tau}^J)$ & $\pi_2 = 1/8$ & D & \wbox & 24.38 & 23.01 & 94.50 & 41.83 & 96.95 & 94.26 & \multicolumn{1}{c}{--} & 80.72 & \textbf{99.13} \\
\wbox & \wbox & \bbox & \wbox & \wbox & Capacity & mL & $\gamma^j$  & $s_3 = \frac{1}{J} \sum_{j=1}^{J} \mathds{1}_j e^{-\varepsilon^j(\gamma^j, \hat{\gamma}^j)}$ & $\pi_3 = 1/8$ & D & \wbox & 24.58 & 40.73 & \multicolumn{1}{c}{--} & 62.57 & 54.79 & 60.57 & \multicolumn{1}{c}{--} & \textbf{72.26} & 59.51 \\
\wbox & \wbox & \wbox & \bbox & \wbox & Container mass & g & $m_c^j$  & $s_4 =  \frac{1}{J} \sum_{j=1}^{J} \mathds{1}_j e^{-\varepsilon^j(m_c^j, \hat{m}_c^j)}$ & $\pi_4 = 1/8$ & D & \wbox & 29.42 & 22.06 & \multicolumn{1}{c}{--} & \multicolumn{1}{c}{--} & \multicolumn{1}{c}{--} & \multicolumn{1}{c}{--} & 49.64 & 40.19 & \textbf{58.78} \\
\wbox & \wbox & \wbox & \wbox & \bbox & Width at top & mm & $w_t^j$  & $s_5 = \frac{1}{J}\sum_{j=1}^{J}{\mathds{1}_j \sigma_1(w_t^j, \hat{w_t}^j)}$ & $\pi_5 = 1/24$ & D & \wbox & 32.33 & 76.89 & \multicolumn{1}{c}{--} & \multicolumn{1}{c}{--} & \multicolumn{1}{c}{--} & \multicolumn{1}{c}{--} & \multicolumn{1}{c}{--} & 69.09 & \textbf{80.01} \\
\wbox & \wbox & \wbox & \wbox & \bbox & Width at bottom & mm & $w_b^j$  & $s_6 = \frac{1}{J}\sum_{j=1}^{J}{\mathds{1}_j \sigma_1(w_b^j, \hat{w_b}^j)}$ & $\pi_6 = 1/24$ & D & \wbox & 25.36 & 58.19 & \multicolumn{1}{c}{--} & \multicolumn{1}{c}{--}  & \multicolumn{1}{c}{--} & \multicolumn{1}{c}{--} & \multicolumn{1}{c}{--} & 59.74 & \textbf{76.09} \\
\wbox & \wbox & \wbox & \wbox & \bbox & Height & mm & $h^j$ & $s_7 = \frac{1}{J}\sum_{j=1}^{J}{ \mathds{1}_j \sigma_1(h^j, \hat{h}^j)}$ & $\pi_7 = 1/24$ & D & \wbox & 42.48 & 64.32 & \multicolumn{1}{c}{--} & \multicolumn{1}{c}{--}  & \multicolumn{1}{c}{--} & \multicolumn{1}{c}{--} & \multicolumn{1}{c}{--} & 70.07 & \textbf{74.33} \\ 
\bbox & \bbox & \bbox & \wbox & \wbox & Filling mass & g & $m_f^j$  & $s_8 = \frac{1}{J} \sum_{j=1}^{J} \mathds{1}_j e^{-\epsilon^j(m_f^j, \hat{m}_f^j)}$ & $\pi_8 = 1/8$* & I & \wbox & 35.06 & 42.31 & 25.07 & 53.47 & 62.16 & 65.06 & \multicolumn{1}{c}{--} & \textbf{70.50} & 65.25 \\
\bbox & \bbox & \bbox & \bbox & \bbox & Object mass  & g & $m^j$  & $s_9 = \frac{1}{J}\sum_{j=1}^{J}{\mathds{1}_j \psi^j(m^j, \hat{F}^j)}$ & $\pi_9 = 1/8$* & I & \bbox & 56.31 & 58.30 & 55.22 & 64.13 & 66.84 & 65.04 & 53.54 & 60.41 & \textbf{71.19} \\
\bbox & \bbox & \bbox & \bbox & \bbox & Pose at delivery & (mm, $^\circ$) & ($\alpha^j$,$\beta^j$)  & $s_{10} = \frac{1}{J}\sum_{j=1}^{J}{\Delta_j(\alpha^j,\beta^j,\eta,\phi)}$ & $\pi_{10} = 1/8$* & I & \bbox & 72.11 & 70.01 & 73.94 & 78.76 & 72.91 & \textbf{80.40} & 60.54 & 73.17 & 79.32 \\
\midrule
\bbox & \bbox & \wbox & \wbox & \wbox & \multicolumn{3}{l}{Joint filling type and level}  & $s_{11} = \bar{F}_1(\lambda^1, \tau^1, \ldots, \hat{\lambda}^1, \hat{\tau}^1, \ldots)$ & \multicolumn{1}{c}{--} & D & \wbox & 10.49 & 8.88 & 77.15 & 24.32 & 77.81 & 76.45 & \multicolumn{1}{c}{--} & 59.32 & \textbf{78.16} \\
\wbox & \wbox & \bbox & \wbox & \bbox & \multicolumn{3}{l}{Container capacity and dimensions} & $s_{12} = {s_3}/{2} + (s4 + s5 + s6)/{6}$ & \multicolumn{1}{c}{--} & D & \wbox & 28.99 & 53.60 & \multicolumn{1}{c}{--} & 31.28 & 27.39 & 30.28 & \multicolumn{1}{c}{--} & \textbf{69.28}  & 68.16 \\
\specialrule{1.2pt}{3pt}{1pt}
\bbox & \bbox & \bbox & \bbox & \bbox & Overall score & & & $S = \sum_{l=1}^{10} \pi_l s_l$ & \multicolumn{1}{c}{--} & I & \multicolumn{1}{c}{--} & 39.11 & 44.51 & 31.52 & 35.89 & 47.04 & 48.35 & 9.05 & 66.16 & \textbf{73.43} \\
\specialrule{1.2pt}{3pt}{1pt}
\multicolumn{21}{l}{\scriptsize{Best performing results for each row highlighted in bold. Results of tasks not addressed shown with a hyphen (--).}}\\
\multicolumn{21}{l}{\scriptsize{For $s_9$ and $s_{10}$, configurations with failures in grasping and/or delivering the containers in simulation using true physical properties as input are annotated and discarded.}}\\
\multicolumn{21}{l}{\scriptsize{For fairness, the residual between 100 and the scores obtained with true measures of the physical properties are added to $s_9$ and $s_{10}$ to remove the impact of the simulator.}}\\
\multicolumn{21}{l}{\scriptsize{KEY -- T:~task, D:~direct score, I:~indirect score, R2S:~measured in the real-to-simulation framework, RAN:~random estimation, AVG:~average from the training set.}}\\
\multicolumn{21}{l}{\scriptsize{* weighted by the number of performed tasks.}}\\
\end{tabular}
\label{tab:scores}
\vspace{-10pt}
\end{table*}
% ===========================

\textbf{Object safety and accuracy of delivery.}
Object safety is the probability that the force applied by the robot, $\tilde{F}$, enables a gripper to hold the container without dropping it or breaking it~\cite{Pang2021ROMAN}. We approximate the force required to hold the container as
\begin{equation}
    \hat{F} \approx \frac{\hat{m} (g + a_{max})}{\mu},
    \label{equ:grasp_force_theoretical}
\end{equation}
where $\hat{m}$ is the annotated object mass; $g=9.81$~m/s$^2$ is the gravitational earth acceleration; $a_{max}$ is the maximum acceleration of the robot arm when carrying the object; and $\mu$ is the coefficient of friction between the container and the gripper ($\mu=1.0$~\cite{Pang2021ROMAN}).
The value of the force applied by the robot to grasp the object is calculated with Eq.~\ref{equ:grasp_force_theoretical} using the predicted object mass $\tilde{m}$. We compute object safety as an exponential function that accounts for the difference between the applied normal force $\tilde{F}^j$ (measured in simulation) and the required normal force, $\hat{F}^j$: 
\begin{equation}
    \psi^j = e^{\frac{| \tilde{F}^j - \hat{F}^j |}{\hat{F}^j} \ln{(1-c)}} = {\ln{(1-c)}}^{\frac{| \tilde{F}^j - \hat{F}^j |}{\hat{F}^j}},
    \label{eq:forcesafety}
\end{equation}
where the normal force is the component of the contact force perpendicular to the contact surface and $c$ controls the sensitivity of $\psi^j$~\cite{Pang2021ROMAN}. A negative difference represents a higher probability of dropping the container, and a positive difference represents a higher probability of breaking the container.

We quantify the accuracy of delivering the container upright and within the target area as
\begin{equation}
    \Delta_j =
    \begin{cases} 
    1 - \frac{\alpha}{\eta} & \text{if } (\alpha < \eta) \text{ and } (\beta < \phi), \\
    0 & \text{otherwise}, \\ 
    \end{cases}
\end{equation}
where $\alpha$ is the distance from the centre of the base of the container to the target location $\boldsymbol{d}$; $\eta$ is the maximum distance allowed from the delivery target location; $\beta$ is the angle between the vertical axis of the container and the vertical axis of the world coordinate system; and $\phi$ is the value of $\beta$ at which the container would tip over. 

We compute the score for object safety, $s_9$, as
\begin{equation}
    s_{9} =  \frac{1}{J}\sum_{j=1}^{J}{ \mathds{1}_j \psi^j(m^j, \hat{F}^j)},
    \label{eq:totobjsafetyscore}
\end{equation}
where the value of the indicator function, $\mathds{1}_j$, is 0 only when either the filling mass or the containers mass is not estimated for each configuration $j$; 
and the score for the delivery accuracy, $s_{10}$, as
\begin{equation}
    s_{10} =  \frac{1}{J}\sum_{j=1}^{J}{ \Delta_j(\alpha^j,\beta^j,\eta,\phi)}.
    \label{eq:deliveryscore}
\end{equation}

The scores $s_9$ and $s_{10}$ are partially influenced by the simulator conditions (e.g, friction, contact, robot control), but we aimed at making the simulated handover reproducible across different algorithms through the annotated object trajectory, starting handover frame, and reconstructed 3D model. 

{\bf Group tasks and overall score.} For joint filling type and level classification ($s_{11}$), estimations and annotations of both filling type and level are combined in $K=7$ feasible classes, and $\bar{F}_1$ is recomputed based on these classes. For joint container capacity and dimensions estimation, we compute the following weighted average:
\begin{equation}
s_{12} = \frac{s_3}{2} + \frac{s4 + s5 + s6}{6}.
\end{equation}

Finally, the overall score is computed as the weighted average of the scores from $s_1$ to $s_{10}$. Note that $s_8$, $s_9$, and $s_{10}$ may use random estimations for either of the tasks not addressed by an algorithm.

%%%%%%%%%%%%%%%%%%%%%%%%%%%%%%%%%%%%%
\subsection{IEEE ICASSP 2022 Challenge entries}

Nine teams registered for the IEEE ICASSP 2022 challenge; three algorithms were submitted for container mass estimation (T4), two algorithms were submitted for classifying the filling level (T1) and type (T2), and  two other algorithms were submitted for estimating the container properties (T3, T4, T5) by three teams. We refer to the submissions of the three teams as A1~\cite{Apicella_GC_ICASSP22}, A2~\cite{Matsubara_GC_ICASSP22}, and A3~\cite{Wang_GC_ICASSP22}. 

A1 solved only the task of container mass estimation (T4) using RGB-D data from the fixed frontal view and by regressing the mass with a shallow Convolutional Neural Network (CNN)~\cite{Christmann2020NTNU}. To increase the accuracy, A1 extracted a set of patches of the detected container from automatically selected frames in a video, and averaged their predicted masses. 
To classify the filling level (T1) and type (T2), A2 used Vision Transformers~\cite{Dosovitskiy2021ICLR}, whereas A3 used pre-trained CNNs (e.g., Mobilenets~\cite{Howard2017Arxiv}) combined with Long Short-Term Memory units or majority voting~\cite{Hochreiter1997LSTM}. Only audio or audio with visual information (RGB) from the fixed, frontal view is preferred as input. To estimate the container properties (T3, T4, T5), A2 used RGB data from the three fixed views, and A3 used RGB-D data from the fixed frontal view. 
A2 used a modified multi-view geometric approach that iteratively fits a hypothetical 3D model~\cite{Xompero2020ICASSP_LoDE}.
A3 fine-tunes multiple Mobilenets via transfer learning from the task of dimensions estimation (T5) to the tasks of container capacity (T3) and mass (T4) estimation~\cite{Wang_GC_ICASSP22}.
These Mobilenets regress the properties using patches extracted from automatically selected frames where the container is mostly visible~\cite{Wang_GC_ICASSP22}. To overcome over-fitting of the limited training data and improve generalisation on novel containers, these Mobilenets are fine-tuned with geometric-based augmentations and variance evaluation~\cite{Wang_GC_ICASSP22}. Overall, A3 is designed to process a continuous stream (online), thus being more suitable for human-to-robot handovers.

Table~\ref{tab:scores} shows the scores of the submissions on the combined CCM test sets. As reference, we provide the results for the properties estimated by a pseudo-random generator (RAN),  by using the average (AVG) of the training set for container capacity, mass, and dimensions; or by the algorithms of four earlier entries to the challenge~\cite{Donaher2021EUSIPCO_ACC,Liu2020ICPR,Ishikawa2020ICPR,Iashin2020ICPR}. A3 achieves the highest $\bar{F}_1$ for filling type classification ($s_2 = 99.13$), and joint filling type and level classification ($s_{11} = 78.16$). A3 is also the most accurate in estimating the container mass ($s_4 = 58.78$), followed by A1 ($s_4=49.64$), and the container dimensions. 
A2 is the most accurate in estimating the  capacity ($s_3 = 72.26$). A2 is also the most accurate for filling mass ($s_8 = 70.50$). A3 has a high accuracy for filling level and type classification, but is affected by its lower accuracy
for capacity estimation.
Among the entries of the challenge at IEEE ICASSP 2022, A3 achieves the best score for object safety ($s_9 = 71.19$) and delivery accuracy ($s_{10} = 79.32$). In conclusion, A3 reaches the highest overall score ($S = 73.43$), followed by A2 ($S = 66.16$).

\section{Conclusion}
\label{sec:conclusion}

Recent, fast advances in machine learning and artificial intelligence have created an expectation on the ability of robots to seamlessly operate in the real world by accurately and robustly perceiving and understanding dynamic environments, including the actions and intentions of humans. 
However, several challenges in audio-visual perception and modelling humans with their hard-to-predict behaviours hamper the deployment of robots in real-world scenarios.

We presented the tasks, the real-to-simulation framework, the scores and the entries to the CORSMAL  challenge at IEEE ICASSP 2022 . 
These new entries complement the algorithms previously submitted to the challenge~\cite{Donaher2021EUSIPCO_ACC,Liu2020ICPR,Ishikawa2020ICPR,Iashin2020ICPR}.

\bibliographystyle{IEEEbib}
\bibliography{xompero}

\begin{thebibliography}{10}

\bibitem{Sanchez-Matilla2020}
R.~{Sanchez-Matilla}, K.~{Chatzilygeroudis}, A.~{Modas}, N.~{Ferreira Duarte},
  A.~{Xompero}, P.~{Frossard}, A.~{Billard}, and A.~{Cavallaro},
\newblock ``Benchmark for human-to-robot handovers of unseen containers with
  unknown filling,''
\newblock {\em IEEE Robotics Autom. Lett.}, vol. 5, no. 2, Apr. 2020.

\bibitem{Medina2016}
J.~R. {Medina}, F.~{Duvallet}, M.~{Karnam}, and A.~{Billard},
\newblock ``A human-inspired controller for fluid human-robot handovers,''
\newblock in {\em Proc. IEEE-RAS Int. Conf. Humanoid Robots}, Cancun, Mexico,
  15--17~Nov. 2016.

\bibitem{Rosenberger2021RAL}
P.~Rosenberger, A.~Cosgun, R.~Newbury, J.~Kwan, V.~Ortenzi, P.~Corke, and
  M.~Grafinger,
\newblock ``Object-independent human-to-robot handovers using real time robotic
  vision,''
\newblock {\em IEEE Robotics Autom. Lett.}, vol. 6, no. 1, pp. 17--23, Jan.
  2021.

\bibitem{Ortenzi2021TRO}
V.~Ortenzi, A.~Cosgun, T.~Pardi, W.~P. Chan, E.~Croft, and D.~Kulić,
\newblock ``Object handovers: A review for robotics,''
\newblock {\em IEEE Trans. Robotics}, vol. 37, no. 6, pp. 1855--1873, Dec.
  2021.

\bibitem{Yang2021ICRA}
W.~Yang, C.~Paxton, A.~Mousavian, Y.~Chao, M.~Cakmak, and D.~Fox,
\newblock ``Reactive human-to-robot handovers of arbitrary objects,''
\newblock in {\em Proc. IEEE Int. Conf. Robotics Autom.}, Xi'an, China,
  30~May--5~June 2021.

\bibitem{Liang2020MultimodalPouring}
H.~Liang, C.~Zhou, S.~Li, X.~Ma, N.~Hendrich, T.~Gerkmann, F-C. Sun, and
  J.~Zhang,
\newblock ``Robust robotic pouring using audition and haptics,''
\newblock in {\em Proc. IEEE Int. Conf. Intell. Robot Syst.}, Las Vegas, NV,
  USA, 24~Oct.~2020--24~Jan. 2021.

\bibitem{Modas2021ArXiv}
A.~Modas, A.~Xompero, R.~Sanchez-Matilla, P.~Frossard, and A.~Cavallaro,
\newblock ``Improving filling level classification with adversarial training,''
\newblock in {\em Proc. IEEE Int. Conf. Image Process.}, Anchorage, Alaska,
  USA, 19--22~Sept. 2021.

\bibitem{Xompero2021_ArXiv}
A.~Xompero, S.~Donaher, V.~Iashin, F.~Palermo, G.~Solak, C.~Coppola,
  R.~Ishikawa, Y.~Nagao, R.~Hachiuma, Q.~Liu, F.~Feng, C.~Lan, R.~H.~M. Chan,
  G.~Christmann, J.-T. Song, G.~Neeharika, C.~K.~T. Reddy, D.~Jain, B.~U.
  Rehman, and A.~Cavallaro,
\newblock ``The {CORSMAL} benchmark for the prediction of the properties of
  containers,'' arXiv:2107.12719v2 [cs.MM], 2021.

\bibitem{Mottaghi2017ICCV}
R.~Mottaghi, C.~Schenck, D.~Fox, and A.~Farhadi,
\newblock ``See the glass half full: Reasoning about liquid containers, their
  volume and content,''
\newblock in {\em Proc. IEEE Int. Conf. Comput. Vis.}, Venice, Italy,
  22--29~Oct. 2017.

\bibitem{Duarte2020ICDL_EpiRob}
N.~F. Duarte, K.~Chatzilygeroudis, J.~Santos-Victor, and A.~Billard,
\newblock ``From human action understanding to robot action execution: how the
  physical properties of handled objects modulate non-verbal cues,''
\newblock in {\em Joint IEEE Int. Conf. Development and Learning and Epigenetic
  Robotics}, Virtual, 26--30~Oct. 2020.

\bibitem{Pang2021ROMAN}
Y.~L. Pang, A.~Xompero, C.~Oh, and A.~Cavallaro,
\newblock ``Towards safe human-to-robot handovers of unknown containers,''
\newblock in {\em IEEE Int. Conf. Robot and Human Interactive Comm.}, Virtual,
  8--12~Aug. 2021.

\bibitem{Xompero_CCM}
A.~Xompero, R.~Sanchez-Matilla, R.~Mazzon, and A.~Cavallaro,
\newblock ``{CORSMAL Containers Manipulation},'' 2020,
\newblock (1.0) [Data set]. Queen Mary University of London.
  \url{https://doi.org/10.17636/101CORSMAL1}.

\bibitem{coumans2019pybullet}
E.~Coumans and Y.~Bai,
\newblock ``{PyBullet}, a {Python} module for physics simulation for games,
  robotics and machine learning,'' \url{http://pybullet.org}, 2016--2019.

\bibitem{Donaher2021EUSIPCO_ACC}
S.~Donaher, A.~Xompero, and A.~Cavallaro,
\newblock ``Audio classification of the content of food containers and drinking
  glasses,''
\newblock in {\em Europ. Signal Process. Conf.}, Virtual, 23--27~Aug. 2021.

\bibitem{Liu2020ICPR}
Q.~Liu, F.~Feng, C.~Lan, and R.~H.~M. Chan,
\newblock ``{VA2Mass}: Towards the fluid filling mass estimation via
  integration of vision \& audio learning,''
\newblock in {\em Proc. IEEE Conf. Pattern Recognit. Workshops and Challenges},
  Virtual, 10--15~Jan. 2021.

\bibitem{Ishikawa2020ICPR}
R.~Ishikawa, Y.~Nagao, R.~Hachiuma, and H.~Saito,
\newblock ``Audio-visual hybrid approach for filling mass estimation,''
\newblock in {\em Proc. IEEE Conf. Pattern Recognit. Workshops and Challenges},
  Virtual, 10--15~Jan. 2021.

\bibitem{Iashin2020ICPR}
V.~Iashin, F.~Palermo, G.~Solak, and C.~Coppola,
\newblock ``Top-1 {CORSMAL} challenge 2020 submission: Filling mass estimation
  using multi-modal observations of human-robot handovers,''
\newblock in {\em Proc. IEEE Conf. Pattern Recognit. Workshops and Challenges},
  Virtual, 10--15~Jan. 2021.

\bibitem{Apicella_GC_ICASSP22}
T.~Apicella, G.~Slavic, R.~Ragusa, P.~Gastaldo, and L.~Marcenaro,
\newblock ``{Container localisation and mass estimation with an RGB-D
  camera},''
\newblock in {\em Proc. IEEE Int. Conf. Acoustics, Speech and, Signal Process.,
  Grand Challenges: Audio-Visual Object Classification For Human-Robot
  Collaboration}, Singapore, 22--27~May 2022.

\bibitem{Matsubara_GC_ICASSP22}
T.~Matsubara, S.~Otsuki, Y.~Wada, H.~Matsuo, T.~Komatsu, Y.~Iioka, K.~Sugiura,
  and H.~Saito,
\newblock ``{Shared transformer encoder with mask-based 3D model estimation for
  container mass estimation},''
\newblock in {\em Proc. IEEE Int. Conf. Acoustics, Speech and, Signal Process.,
  Grand Challenges: Audio-Visual Object Classification For Human-Robot
  Collaboration}, Singapore, 22--27~May 2022.

\bibitem{Wang_GC_ICASSP22}
H.~Wang, C.~Zhu, Z.~Ma, and C.~Oh,
\newblock ``Improving generalization of deep networks for estimating physical
  properties of containers and fillings,''
\newblock in {\em Proc. IEEE Int. Conf. Acoustics, Speech and, Signal Process.,
  Grand Challenges: Audio-Visual Object Classification For Human-Robot
  Collaboration}, Singapore, 22--27~May 2022.

\bibitem{Christmann2020NTNU}
G.~Christmann and J.~Song,
\newblock ``{2020 CORSMAL Challenge - Team NTNU-ERCReport},'' 2020,
\newblock
  \url{https://corsmal.eecs.qmul.ac.uk/resources/challenge/2020.11.30_CORSMAL_NTNU-ERC_Report.pdf}.

\bibitem{Dosovitskiy2021ICLR}
A.~Dosovitskiy, L.~Beyer, A.~Kolesnikov, D.~Weissenborn, X.~Zhai,
  T.~Unterthiner, M.~Dehghani, M.~Minderer, G.~Heigold, S.~Gelly, J.~Uszkoreit,
  and N.~Houlsb,
\newblock ``An image is worth 16x16 words: transformers for image recognition
  at scale,''
\newblock in {\em Int. Conf. Learning Rep.}, Virtual, 3--7~May 2021.

\bibitem{Howard2017Arxiv}
A.~G. Howard, M.~Zhu, B.~Chen, D.~Kalenichenko, W.~Wang, T.~Weyand,
  M.~Andreetto, and H.~Adam,
\newblock ``Mobilenets: Efficient convolutional neural networks for mobile
  vision applications,'' arXiv:1704.04861, 2017.

\bibitem{Hochreiter1997LSTM}
S.~Hochreiter and J.~Schmidhuber,
\newblock ``{Long Short-Term Memory},''
\newblock {\em Neural Computation}, vol. 9, no. 8, pp. 1735--1780, 11 1997.

\bibitem{Xompero2020ICASSP_LoDE}
A.~Xompero, R.~Sanchez-Matilla, A.~Modas, P.~Frossard, and A.~Cavallaro,
\newblock ``Multi-view shape estimation of transparent containers,''
\newblock in {\em Proc. IEEE Int. Conf. Acoustics, Speech and Signal Process.},
  Barcelona, Spain, 4-8~May 2020.

\end{thebibliography}

\end{document}